\documentclass[conference]{IEEEtran}
\usepackage{cite}
\usepackage{amsmath,amssymb,amsfonts,commath}
\usepackage{algorithmic}
\usepackage{graphicx}
\usepackage{textcomp}
\usepackage{tabularx}
\usepackage{xcolor}
\usepackage{caption}
\usepackage{comment}
\usepackage{subcaption}
\usepackage{todonotes}
\usepackage{diagbox}
\usepackage{amsthm}
\usepackage{url}

\usepackage{glossaries}
\usepackage{glossaries-extra}

\usepackage{nomencl}
\makenomenclature

\usepackage{etoolbox}
\renewcommand\nomgroup[1]{%
  \item[\bfseries
  \ifstrequal{#1}{M}{Symbols}{%
  \ifstrequal{#1}{S}{Sets}{%
  \ifstrequal{#1}{P}{Superscript}{%
  \ifstrequal{#1}{I}{Indices}{%
  \ifstrequal{#1}{T}{Parameters}{%
  \ifstrequal{#1}{V}{Variables}{}}}}}}%
]}

\newcommand{\lDisco[1]}{$\setminus {#1}$}
\newcommand{\lReco[1]}{$={#1}$}
\newcommand{\compo}{\odot}
\newcommand{\RecoSet}{$R$}

\newcommand{\DiscoSet}{$O$}

\newcommand{\antoine}[1]{}
\newcommand\Tau{\mathrm{\textit{T}}}

\def\BibTeX{{\rm B\kern-.05em{\sc i\kern-.025em b}\kern-.08em
    T\kern-.1667em\lower.7ex\hbox{E}\kern-.125emX}}
\begin{document}

\title{Superposition theorem for flexible grids}

\author{Antoine Marot, Noureddine Henka, Benjamin Donnot, Sami Tazi, IEEE members}
\author{\IEEEauthorblockN{Antoine Marot \& Benjamin Donnot}\\
\IEEEauthorblockA{\textit{AI Lab}\\
\textit{RTE France, Paris}
}
\and
\IEEEauthorblockN{Noureddine Henka \& Sami Tazi}\\
\IEEEauthorblockA{\textit{R\&D Department}} 
\textit{RTE France,Paris}
}
\markboth{Journal of \LaTeX\ Class Files,~Vol.~14, No.~8, August~2021}%
{Shell \MakeLowercase{\textit{et al.}}: A Sample Article Using IEEEtran.cls for IEEE Journals}
\maketitle

\begin{abstract}
Flexible grid topology has become a key enabler of flexibility in modern power grids, particularly for congestion management. Studying the effects of combinatorial topological changes is therefore of significant interest, though it remains computationally intensive in most cases. To address this, we revisit the superposition theorem, which has served as the foundation for the decomposition of numerous power system problems over the past decades, particularly those involving changes in generation and loads. However, its application has traditionally been restricted to fixed grid topologies, breaking down as soon as a topology change occurs.
In this paper, we extend the superposition theorem to accommodate varying grid topologies by leveraging well-known distribution factors. This unified framework applies to all types of topological changes, including line disconnection and reconnection, as well as bus splitting and merging. We provide numerical experiments to validate our approach and highlight its advantages in terms of speed-up and interpretability. Finally, we demonstrate its application in two key use cases: remedial action search and updated security analysis following a topological change. Our results show that the proposed approach achieves more than an order-of-magnitude speed-up for real-sized grids compared to the fastest power flow solvers.
\end{abstract}


\nomenclature[M]{\(\lDisco[l]\)}{Line $l$ disconnected}
\nomenclature[M]{\(\lReco[l]\)}{Line $l$ reconnected}
\nomenclature[M]{\(\compo\)}{Composition symbol of topologies}
\nomenclature[P]{\(ref\)}{Reference variable state to start from }
\nomenclature[P]{\(tgt\)}{Target variable state reached after changes }

\nomenclature[I]{\(i\)}{Unitary change in topology}
\nomenclature[I]{\(k\)}{Unitary change in net nodal power balance}
\nomenclature[I]{\(l\)}{Impacted line}
\nomenclature[I]{\(o\)}{Outage line}
\nomenclature[I]{\(r\)}{Reconnected line}

\nomenclature[S]{\(\DiscoSet\)}{Set of lines $l$ getting disconnected in outage}
\nomenclature[S]{\(\RecoSet\)}{Set of lines $l$ getting reconnected}
\nomenclature[S]{\(\RecoSet_{\setminus o}\)}{Set of lines $l$ getting reconnected except line $o$}

\nomenclature[V]{\(\Tau\)}{A grid topology}
\nomenclature[V]{\(\Tau^{ref}\)}{A reference topology to start from}
\nomenclature[V]{\(\Tau^{tgt}\)}{A target topology reached after $N_{\tau}$ topological changes such that $\Tau^{tgt}=\Tau^{ref} \odot \tau_1 \odot ...\odot \tau_{n_\tau}$}
\nomenclature[V]{\(\Tau_{O}\)}{A topology with $l \in O$ lines disconnected}
\nomenclature[V]{\(\Tau_{R}\)}{A topology with $l \in R$ lines reconnected}
\nomenclature[V]{\(\tau_i\)}{$i^{th}$ unitary topology change on a line or at a substation}
\nomenclature[V]{\(\tau_{\setminus l}\)}{Unitary line $l$ disconnection}
\nomenclature[V]{\(\tau_{=l}\)}{Unitary line $l$ reconnection}
\nomenclature[V]{\(\Tau^{\tau_i}\)}{Grid topology $\Tau^{ref}\odot\tau_i$ after applying unitary change $\tau_i$ to \Tau^{ref}}
\nomenclature[V]{\(P_{g-d}\)}{Net nodal power balance vector of dim $n_{bus}$, with generations (g) minus distribution loads (d) at nodes}
\nomenclature[V]{\(P_{g-d}^{'}\)}{$P_{g-d}$ augmented by $2n_O$ null elements to match the dimension of $Cf^{\Tau_O}$ in the equivalent canceling flow model}
\nomenclature[V]{\(\delta P_{g-d}^k\)}{$k^{th}$ unitary change in net nodal power balance vector}
\nomenclature[V]{\(PF(P_{g-d},\Tau)\)}{Lines power flow vector of dim $n_l$ under $P_{g-d}$ nodal power balance and topology $\Tau$ with DC approximation}
\nomenclature[V]{\(PF(T)\)}{Simpler $PF(P_{g-d}^{ref},\Tau)$ notation. \textbf{If nodal power balance variable is not explicit, it is assumed to be $P_{g-d}^{ref}$ in all other definitions}}
\nomenclature[V]{\(pf^{\Tau}_l\)}{Line $l$ power flow from $PF(\Tau)$}
\nomenclature[V]{\(cf^{\Tau}_l\)}{Line $l$ canceling flow value in topology \Tau}
\nomenclature[V]{\(Cf^{\Tau_O}\)}{Canceling flow nodal vector of dim $n_{bus}^{ref} + 2n_O$ under disconnected lines $l \in O$, sum of Cf^{\Tau}_l }
\nomenclature[V]{\(Cf^{\Tau}_{l}\)}{$Cf^{\Tau}$ filtered on line $l$, so with null elements except at virtual buses infinitely close to line $l$ extremities such that $Cf^{\Tau_O}=\sum_{l \in O} Cf^{\Tau_O}_l$}
\nomenclature[V]{\(vf^{\Tau}_l\)}{Line $l$ virtual induced flow in topology \Tau}

\nomenclature[V]{\(\theta^{\Tau}_{l_{or}},\theta^{\Tau}_{l_{ex}}\)}{Voltage angle at each line $l$ node (origin and extremity) in topology \Tau}

\nomenclature[V]{\(
\Delta\theta^{\Tau}_l\)}{Difference in line $l$ voltage angles in topology \Tau}

\nomenclature[V]{\(
LODF_{o}\)}{Line outage $o$  distribution factors vector \\ & on all lines $n_l$ in \Tau^{ref}}

\nomenclature[V]{\(
lodf_{o,l}\)}{Line outage $o$  distribution factor \\ & on line $l$ in \Tau^{ref}}

\nomenclature[V]{\(
lcdf_{r,l}\)}{Line closing $c$  distribution factor \\ & on line $l$ in \Tau^{ref}}

\nomenclature[V]{\(\alpha \text{ and } \beta^i\)}{Superposition theorem scalar coefficients for $PF(T^{ref})$ and  $PF(T^{\tau_i})$ }

\nomenclature[V]{\(\alpha_R,\beta^i_R (\text{resp. } \alpha_O,\beta^i_O\))}{Superposition theorem coefficients for line reconnections (resp. line disconnections)}

\nomenclature[V]{\(TSF^{\tau_i}\)}{Vector of Topology Sensitivity Factor on all lines after applying $\tau_i$ in \Tau^{ref}}

\nomenclature[V]{\(tsf^{\tau_i}_{l}\)}{ $TSF^{\tau_i}$ on real or virtual flow of line $l$}

\nomenclature[T]{\(n_\tau, n_O, n_R \)}{number of considered unitary topology changes (resp. line disconnections, line reconnections) }
\nomenclature[T]{\(n_l\)}{number of lines on the grid}
\nomenclature[T]{\(n_{bus}\)}{number of nodes in $T^{ref}$}
\nomenclature[T]{\(\sigma_l \)}{Conductance of line l, inverse of impedance x_l}

\renewcommand{\nomname}{Glossary}

\section{Introduction}
To succeed at the energy system decarbonization and mitigate the effect of climate change, the largest emitting countries are undergoing a steep Energy Transition worldwide. Power systems are at the forefront of this transition,  responsible for integrating a high share of renewable energy but also enabling greater electrification of usages, while accommodating for better efficiencies. For existing power grids, we are moving from a mostly predictable environment with known demand patterns, controllable generation, and middle life reliable infrastructure assets to an environment with a lot more uncertainties given intermittent generation, aging infrastructure requiring more outage management, active demand, and numerous market players \cite{marot2022perspectives}. In addition, more care must be taken to the development and use of resources given ecological considerations. Hence, more flexible operations with existing infrastructure are needed now. In particular, one could make the grid more flexible by dynamically adapting the grid topology, leveraging this overall but currently underutilized flexibility especially in meshed areas. Indeed, a change in topology changes the line power flows that one wants to monitor or control to avoid congestions. The topology can hence be regarded as an important variable. It can help, in particular, keep the grid safe from congestions while integrating renewables at scale \cite{little2021optimal,numan2020mobilizing,marot2021learning}, avoiding curtailment and redispatching costs \cite{fuxjager2023reinforcement,lehna2023managing,subramanian2021exploring}, increasing market capacity through flow-based domain calculation \cite{aluisio2023remedial}, or facilitating outage management \cite{crognier2021grid}. This is especially true for meshed or interconnected grids for which possible topology configurations are numerous. The latest developments show that topology could eventually be controlled at a complex level using artificial intelligence \cite{dorfer2022power} unlocking even more flexibility.

Considering topology as an active variable was not so obvious in the past. Topology has often been regarded as fixed, besides possible line contingencies that could result in unplanned outages. Indeed, most of the research work up to today has only considered one reference topology. This is also because topology changes introduce costly refactoring or combinatorial computations in classical (optimal) power flow formulations, especially when the admittance matrix dimension changes.

However, there has been a renewed interest in the topology to play a more active role in terms of flexibility with the use of transmission switching \cite{Numan2023,ruiz2016security,dehghanian2017power}, substation reconfiguration \cite{morsy2022security,Ruiz2017,Goldis2017} as well as dealing with more complex planned outage management \cite{crognier2021grid}. This involves combinations of all types of topological changes, not only line disconnections as considered historically \cite{glavitsch1985switching}, but also now active transmission switching including offload or outaged line reconnections, node splitting and merging. Tools with better performance are now needed when considering those variations\cite{marot2020learning,marot2021learning,viebahn2022potential}.

Even if not mainstream and rarely deployed in real processes, transmission switching as an active control means has been around since the 80s in research work with a focus on active line disconnections from a meshed reference grid topology \cite{glavitsch1985switching}. It has primarily been considered in two directions, which we quickly review as building blocks for our work. One has been the definition and use of distribution factors 
for interactive human analysis and operational study as in \cite{Sauer2001, marot2018expert, viebahn2024potential}. The Optimal Power Flow (OPF) formulation, and more specifically the optimal transmission switching (OTS), has been the complementary research direction \cite{schnyder1990security,hedman2011review,dehghanian2017power, Numan2023}, embedding a more automated process perspective. 

About distribution factors, the first research avenue, its workhorse has been Line Outage Distribution Factors (LODFs) \cite{shahidehpour2003market,wood2013power,bhattacharya2012operation} historically. GLODF were later proposed as a generalization of LODF for multiple outages \cite{guler2007generalized}, relying on Power Transfer Distribution Factors (PTDFs). Early considerations of GLODF can already be found in \cite{schnyder1990security}. A more comprehensive list of various distribution factors can be found in \cite{Sauer2001}. They define in particular LCDF (Line Closing Distribution Factors) of special interest when considering line reconnections. It was again defined in terms of PTDFs. Lately, Bus Splitting Distribution Factors (BSDFs) has been proposed \cite{van2023bus} to compute unitary bus splitting change from PTDFs. Nevertheless, as the authors noted, a single generic framework is still lacking to consider all those changes and related factors at once, much like GLODF, but combining all types. This is the direction we pursue in this paper, using the Superposition Theorem (ST) as a unifying framework to connect them all, eventually generalizing this century-old theorem for changing topologies. It is noteworthy that the superposition theorem was originally the foundation of GLODF-like derivation from \cite{schnyder1990security} 
before PTDF became the cornerstone of later developments. However, relying on the PTDF matrix, when not readily available, requires the computation of an intermediate step bringing a significant computational overhead at the start for computing new power flows. To directly compute new power flows from existing power flow results without additional steps or data required, we chose to get back to this first and foundational superposition principle at the root of distribution factors to achieve the targeted generalization.

With regard to generalization, the second research avenue based on OPF and OTS has been paving the way towards it by aiming to frame things into a single optimization problem. Usage of mixed topological changes including line disconnections, bus splitting and merging, could already be found in \cite{zaoui2005coupling,fliscounakis2007topology} with the formulation of heterogenous types of inequalities and penalty parameters to calibrate.  Substation reconfigurations were later introduced more thoroughly in \cite{Heidarifar2016} which led \cite{morsy2022security} to derive an equivalent formulation between bus splitting and line switching when augmenting the network model. Finally, \cite{Goldis2017} dealt with all 4 types of topological changes and reformulated OTS with the use of distribution factors, connecting those two aforementioned research directions. They however used a workaround by relying on two complementary formulations, to model things either as all disconnections (Breaker Closing Transactions - BCT) or as all reconnections (Breaker Opening Incremental Flow - BOIF). Hence, the computation does not depart from the current topology. In BCT case, it departs with all switching assets opened, while for BOIF case, it departs with all of them closed. For both cases, this requires a significant pre-computation step of a PTDF-like matrix in the departing topology, different from the current topology at hand. It is also preferred in operations to minimize changes from the current topology and hence explore necessary and sufficient changes departing from it, stopping the search early enough. 

In this paper, we show that one can directly and efficiently compute new power flows in a single linear formulation for all types of combined changes. It effectively starts from the currently observed topology and without intermediate computation steps. For this, we revisit the century-old superposition theorem principle \cite{boylestad2009electronic}. The main contributions of this paper are:
\begin{itemize}
    \item Presenting a unified framework framed as an extended superposition theorem for computing efficiently all combined topological changes in Section \ref{sec:gen_sup_th}   
    \item Proving a new formulation of Generalized Line Closing Distribution Factors in Section \ref{sec:line_reco}
    \item Deriving fast and frugal power flow calculations for flexible grid topology without pre-computing steps or grid information knowledge in Section \ref{sec:power_flow_Rombined}
    \item Demonstrating its applicability and performance on canonical and widely used applications in power grids, and their possible extensions in Section \ref{sec:results}
\end{itemize}

For reading convenience, all the paper notation for the following sections can be found in the \textbf{Glossary} at the end.

\section{Extended superposition theorem framework}
\label{sec:gen_sup_th}
 
We consider a grid with generation g and load distribution d at $n_{bus}$ electrical nodes, represented by the nodal power balance vector $P_{g-d}$. Branches connections at every node represent the grid topology that we denote by $\Tau$. Our aim is to compute power flows $PF(P_{g-d},\Tau)$. In this paper, We restrict ourselves to the DC approximation \cite{stott2009dc} and assume that the grid remains connected into one single component.

\subsection{Standard Superposition Theorem}

Given a reference nodal power balance vector $P_{g-d}^{ref}$ and k additions of generation and load changes  $\delta P_{g-d}^k$ (for instance a change of +10MW at a node and -10MW at another) summing up to target nodal power balance vector $P_{g-d}^{tgt}$,  the standard ST provides a sum decomposition of resulting power flows under the same reference grid topology $\Tau^{ref}$: 


\begin{multline}
\label{eqn:injection_theorem}
        PF(P_{g-d}^{tgt},\Tau^{ref})=PF(P_{g-d}^{ref},\Tau^{ref})+\sum_{k} PF(\delta P_{g-d}^k,\Tau^{ref})\\
        \text{with } P_{g-d}^{tgt}=P_{g-d}^{ref}+\sum_{k} \delta P_{g-d}^k
\end{multline}

Thus, one can infer a target grid state without resolving the power flow equations again if the initial and unitary grid states are known. ST has proven very useful for decomposing the problem analytically, allowing for either more efficient computations or better interpretability when analyzing some grid phenomena. We aim to transpose this handy tool to topological changes.

\subsection{Extended Superposition Theorem}

\begin{figure}[h]
    \centering
    \includegraphics[scale=0.27]{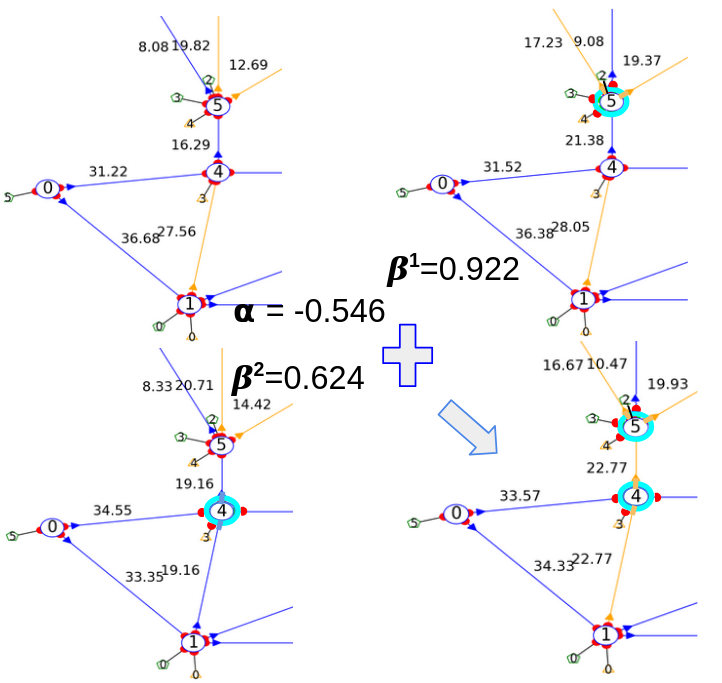}
    \caption{EST example on IEEE14, starting from a meshed topology (top left) to which 2 node splitting actions at substations 4 and 5 are applied (bottom right). Displayed EST coefficients are derived from initial and unitary action states.}
    \label{fig:node_spliiting_example}
\end{figure}

To extend it similarly to topological changes, we start from $\Tau^{ref}$ as the reference topology to which we can apply unitary topological changes $\tau_i$ each resulting in topology $\Tau^{\tau_i}=\Tau^{ref} \odot \tau_i$. When combined in indifferent order, we reach a target topology $\Tau^{tgt}=\Tau^{ref} \odot \tau_1 \odot ...\odot \tau_{n_\tau}$. As illustrated in Figure \ref{fig:node_spliiting_example}, we will prove the following Extended Superposition Theorem (EST): 
\begin{multline}
\label{eqn:extended_ST}
    PF(P_{g-d}^{ref},\Tau^{tgt})=\alpha PF(P_{g-d}^{ref},\Tau^{ref}) + \sum_{i=1}^{n_\tau} \beta^i PF(P_{g-d}^{ref},\Tau^{\tau_i})\\
    \text{with } \alpha=1-\sum_{i=1}^{n_\tau}\beta^i
\end{multline}


Note that this decomposition is a weighted linear combination instead of a pure linear one as before. This requires us to compute these weights. We will demonstrate that finding the $\beta^i$ stems from solving a linear system of dimension the number of considered unitary changes $n_\tau$: 

\begin{gather}
\label{eqn:ST_linear_system}
 \begin{bmatrix} 1 & \text{tsf}^{\Tau^{\tau_2}}_{l_1} & ... & \text{tsf}^{\Tau^{\tau_n}}_{l_1} \\ 
 \text{tsf}^{\Tau^{\tau_1}}_{l_2} & 1 & ... & \text{tsf}^{\Tau^{\tau_n}}_{l_2} \\
  ... & ... & ... & ... \\
   \text{tsf}^{\Tau^{\tau_1}}_{l_n} & \text{tsf}^{\Tau^{\tau_2}}_{l_n} & ... & 1\\
 \end{bmatrix}
  \begin{bmatrix}
    \beta1 \\
    \beta2 \\
    ... \\
    \beta n
  \end{bmatrix}
 =
   \begin{bmatrix}
    1 \\
    1 \\
    ... \\
    1
  \end{bmatrix}
\end{gather}
Where  $\text{tsf}^{T^{\tau_i}}_{l_j}$ are Topology Sensitivity factors (TSF) reflecting the change in power flow of asset $l_j$ after a unitary topology change $\tau_i$ applied on asset $l_i$ from $T^{ref}$:  
\begin{multline}
\label{eqn:TSF}
    \text{tsf}^{T^{\tau_i}}_{l_j}=\frac{x^{T^{ref}}_{l_j}-x^{T^{\tau_i}}_{l_j}}{x^{T^{ref}}_{l_j}}
    \text{with } x_{l_j}=pf_{l_j} \text{ or } \Delta \theta_{l_j} \text{ given } \tau_i 
\end{multline}

$pf_{l_j}$ is the power flowing in asset $l_j$ and $\Delta \theta_{l_j}$ the voltage angle difference at its extremities. Given $\sigma_{l_j}$ this asset's conductivity, they are tied to Ohm's law when connected:
\begin{equation}
    pf_{l_j}=\sigma_{l_j} \Delta \theta_{l_j}
\end{equation}

Yet (\ref{eqn:ST_linear_system}) and (\ref{eqn:TSF}) only rely on the knowledge of reference and unitary topological change power flows. $\beta^i$ coefficients are functions of those states only $\beta^i=f^i(PF(T^{ref}),PF(T^{\tau_1}),...,PF(T^{\tau_{n_\tau}}))$ using simpler notation $PF(T)=PF(P_{g-d}^{ref},\Tau)$. This ensures that equation (\ref{eqn:extended_ST}) is indeed a true superposition of these single states. 
Considering that such measurements for assets switched and of interest solely are available, no other knowledge is needed, such as measurements on other assets, intrinsic grid properties, complete adjacency matrix or full topology information. This will make the computation frugal without resolving again any power flow equations.

\antoine{what happens in the case $x^{T^{ref}}_l$ is null ? in the case we consider an initial null flow for a line we disconnect, it is as if it is already disconnected. Hence, the state will be the same after this unitary action, hence LODF is null and we have one less equation to consider in (15). We have a degenerate system for which we can set beta is null, and this unitary state, which is the same as the original state, will be accounted in the alpha directly. In the case the asset is initially disconnected, connecting the line will not create any additional flow, resulting in a staggering state, and we will have a similar degenerated state for beta}

This linear system form is generic to all four types of topological changes, as we will demonstrate in the next sections. Therefore this makes it a proper candidate for achieving a unified framework for combining them all in a single and direct computation scheme. In the next sections, we provide a step-by-step demonstration of all four types of change. 

\section{EST for line disconnections}
\label{sec:disco_sup_th}
In this section, we (re)-demonstrate with our formalism how this EST applies to line disconnections $o \in O$ in the first place. It relates to well-known LODFs, and could have similarly been based on other well-known PTDFs.  

We first set up the stage for this demonstration and later ones across the paper by revisiting a useful equivalent modeling.

\subsection{Two equivalent models of line disconnections}
\begin{figure}[h]
    \centering
    \includegraphics[scale=0.35]{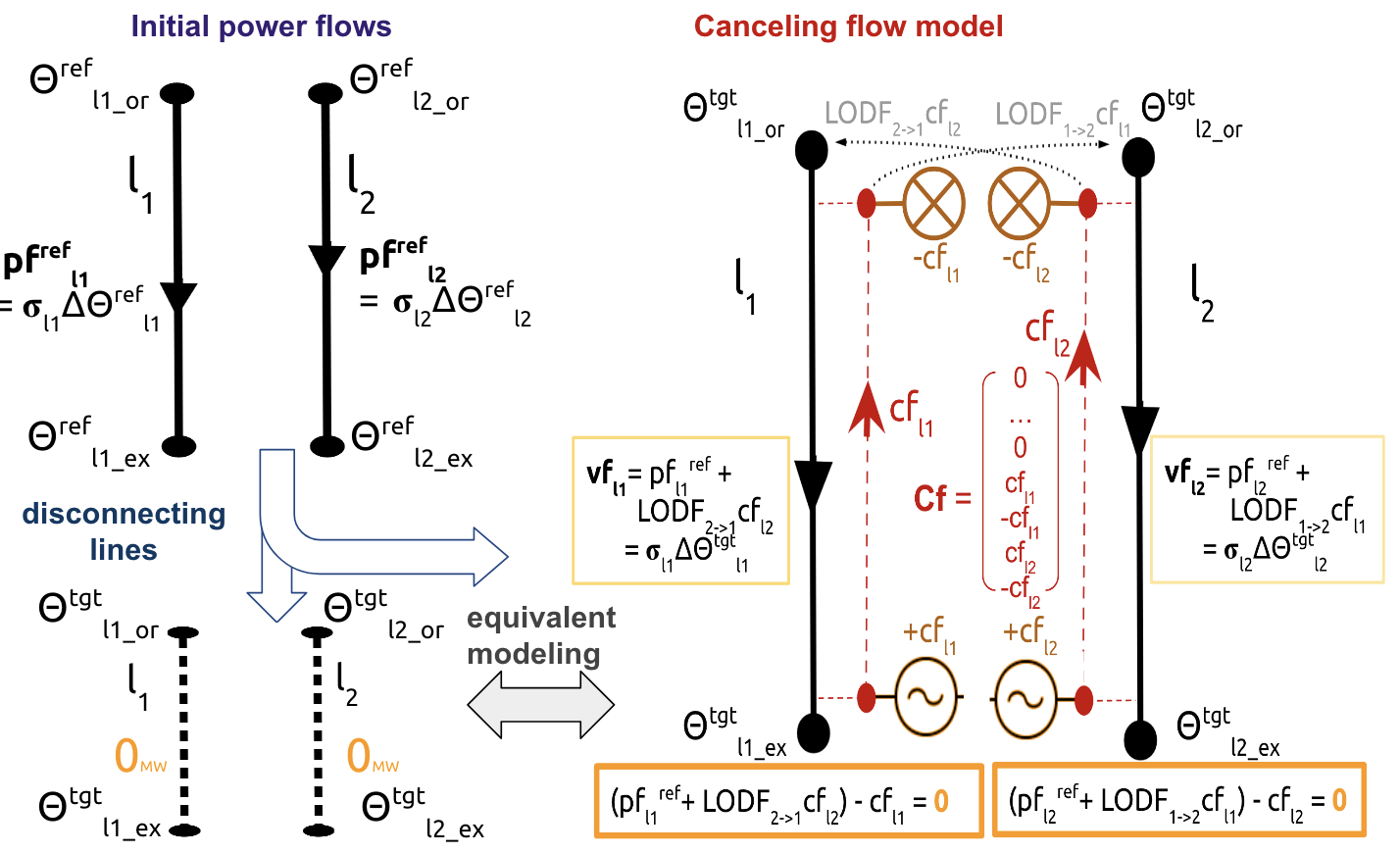}
    \caption{Top left, power Flows in the reference topology for the two lines to disconnect. Right and bottom left, the two lines disconnected in two equivalent models: the standard one with physical line disconnections on top, and the canceling flow model at the bottom. The lines remains virtually connected but with equivalent null flow: the virtual induced flow $vf_l$ gets canceled out by the reversed direction canceling flow $cf_l$.}
    \label{fig:cancelling_flow}

\end{figure}

We now describe an equivalent grid state model for disconnected lines that will be the foundation for EST demonstrations.
Disconnecting a line $o$ resulting in topology $\Tau_o$ is actually equivalent to virtually injecting a canceling flow $cf_o^{\Tau_O}$ along that line $o$ at virtual buses infinitely 
close to the two line extremity buses \cite{guler2007generalized,ruiz2016security}, while keeping this line virtually connected. To inject such a canceling flow, one can add a load and generator of its power value distributed at those two virtual buses. This also generalizes to multiple line disconnections that belong to a set $O$, resulting in topology $\Tau_O$, keeping all of these lines virtually connected but ensuring that their flow is null by injecting canceling flows at these lines
as in Figure \ref{fig:cancelling_flow}. Note that for multiple line disconnections, canceling flows are different from their initial line flows, as influences between the line flows have to be taken into account.   

Given $n_O$ disconnected lines from $\Tau^{ref}$, this canceling flow injections equivalence can be represented by:
\begin{equation}
\label{eqn:virtual_flow_model}
    PF(P_{g-d}^{ref},\Tau_O) = PF({P'}_{g-d}^{ref}+\sum_{o \in O} Cf^{\Tau_O}_o,\Tau^{ref})
\end{equation}
$Cf^{\Tau_O}_o$ of dimension $n_{bus} +2n_O$ is here the nodal injection vector form of $cf_o^{\Tau_O}$ with non-null elements of positive and negative $cf_o^{\Tau_O}$ at the virtual buses infinitely close to line $l$ extremities, and null nodal values otherwise. Note that ${P'}_{g-d}^{ref}$ is $P_{g-d}^{ref}$ augmented with $2n_O$ null elements to match $Cf^{\Tau_O}_o$ dimension given $2n_O$ virtual buses to consider. 

The grid topology in the virtual injection model is $\Tau^{ref}$ as the lines remain virtually connected. 
 These models are equivalent, as they result in the same power flows. Given Ohm's law, this also implies that voltage angles at nodes also remain the same under the same reference voltage angle node. 
In this equivalent model, there exist in addition a ohm's law induced virtual power flow $vf^{\Tau_O}_o$ at line $o$ still virtually connected:
\begin{equation}
\label{eqn:virtual_induced_flow}
    vf_o^{\Tau_O} =\sigma_o \Delta \theta_o^{\Tau_O}
\end{equation}
As this equivalent model aims at representing the system state following the disconnections of lines $o \in O$, the resulting virtual flows $vf^{\Tau_O}_o$ 
should be exactly compensated by their canceling flows as in Figure \ref{fig:cancelling_flow}: 
\begin{equation}
\label{eqn:canceling_flow_equation}
    vf_o^{\Tau_O} - cf_o^{\Tau_O} = 0 \text{ for all } l \in O
\end{equation}


This equivalent modeling is of interest since one can apply standard ST (\ref{eqn:injection_theorem}) to (\ref{eqn:virtual_flow_model}) such that:
\begin{equation}
\label{eqn:virtual_flow_model_st}
    PF(P_{g-d}^{ref},\Tau_O) = PF(P_{g-d}^{ref},\Tau^{ref})+\sum_{o \in O} PF(Cf^{\Tau_O}_o,\Tau^{ref})
\end{equation}

This is the foundational step for further deriving this EST for topological changes as we will see. However, we will need to derive $Cf^{\Tau_O}_o$ for all lines $o \in O$ of interest.

\subsection{Line disconnections demonstration}
\label{sec:line_disco_th}
Let's first consider the restricted topological changes case of line disconnections $o \in O$. We depart from a reference topology in which these lines are connected \textbf{$\Tau^{ref}=\Tau_R$} and reach a topology \textbf{$\Tau^{tgt}=\Tau_O$} with all lines disconnected $o \in O$.
The proposed EST in this case has already been indirectly demonstrated in the Generalize LODF paper \cite{guler2007generalized}. Indeed, they compute multi-outage power flow solving a linear system of LODF, which is composed of reference and unitary outage power flow states.

%
%

To derive the linear system of equations to solve in our formalism, we work with the equivalent virtual injection models. In the one-line $o$ disconnection case, the virtual flow is equivalent to the initial power flow $vf_o^{\Tau^{\tau_{\setminus o}}}=pf_o^{\Tau^{ref}}$. As from (\ref{eqn:canceling_flow_equation}), it is canceled in the reverse direction by a canceling flow of a similar value $cf_o^{\Tau^{\tau_{\setminus o}}} = pf_o^{\Tau^{ref}}$  .

In the multiple line disconnection case, virtually injected canceling flows on other lines to disconnect (but virtually connected in the equivalent model) induce additional virtual flows on the line of interest as from (\ref{eqn:virtual_flow_model_st}):
\begin{equation}
    PF(P_{g-d}^{ref},\Tau^{tgt}) = PF(P_{g-d}^{ref},\Tau^{ref})+\sum_{o \in O} PF(Cf^{\Tau^{tgt}}_o,\Tau^{ref})
\end{equation}

To derrive the new powerflow terms, we use LODFs with effect on line $l$ after line outage $o$:
\begin{equation}
\label{eqn:LODF}
    lodf_{o,l} = \frac{pf_l^{\Tau^{\tau_{\setminus o}}} - pf_l^{\Tau^{ref}}}{pf_o^{\Tau^{ref}}}
\end{equation}
$LODF_o$ is the vector form of outage $o$ effect for all lines $l$.
Note that LODF remain constant for a given topology such as $\Tau^{ref}$. This is also a well-defined term even when $pf_o^{\Tau^{ref}}=0$. In fact, in that case, this is as if the line has already been disconnected from the start. Disconnecting it will not change the flows at all. Hence, the numerator will be null as well, and the resulting LODF will be null too.

In the equivalent modeling, as its topology remains $\Tau^{ref}$, we can hence reuse the same LODF computed in $\Tau^{ref}$.


%



Using $LODF$, $PF(Cf^{\Tau^{tgt}}_o,\Tau^{ref})$ is simply:
\begin{equation}
\label{eqn:lodf_ST_th}
PF(Cf^{\Tau^{tgt}}_o,\Tau^{ref})=LODF_o\times cf_o^{\Tau^{tgt}}
\end{equation}

This eventually leads by substitution to the equation:  
\begin{equation}
\label{eqn:topo_theorem}
    PF(\Tau^{tgt})=PF(\Tau^{ref}) + \sum_{o \in O} LODF_o\times cf_o^{\Tau^{tgt}}
\end{equation}


As $pf_l^{\Tau^{tgt}}=0$ for disconnected lines, this leads to the system of $n_O$ independent canceling flow equations: 

\begin{multline}
\label{eqn:linear_system_line_disconnection}
    pf_l^{\Tau^{ref}}+ \sum_{o \in O } lodf_{o,l} \times  cf_o^{\Tau^{tgt}} =0 \text{ for all } l \in O \\
    pf_l^{\Tau^{ref}} + \sum_{o \in O, o \neq l } lodf_{o,l} \times cf_o^{\Tau^{tgt}}= cf_l^{\Tau^{tgt}} \text{ as $lodf_{l,l}=-1$ }
\end{multline} 

From (\ref{eqn:canceling_flow_equation}) the virtual induced flow $vf_l$ value on line $l$ is then:
\begin{equation}
    vf_l^{\Tau^{tgt}} = cf_l^{\Tau^{tgt}}= pf_l^{\Tau^{ref}} + \sum_{o \in O, o \neq l } lodf_{o,l} \times cf_o^{\Tau^{tgt}}\\
\end{equation}
The other canceling flow contributions are added up to the initial flow $pf_l^{\Tau^{ref}}$ to result in the induced flow. Substituting LODF definition from (\ref{eqn:LODF}) in (\ref{eqn:lodf_ST_th})  the EST is:
\begin{multline}
\label{eqn:sup_topo_theorem}
\begin{split}
    PF(\Tau^{tgt}) & =PF(\Tau^{ref}) + \sum_{o \in O} (PF(\Tau^{\tau_{\setminus o}}) - PF(\Tau^{ref})) \frac{cf_o^{\Tau^{tgt}}}{pf_o^{\Tau^{ref}}} \\
    & = \alpha \times PF(\Tau^{ref}) +  \sum_{o \in O} \beta^o \times PF(\Tau^{\tau_{\setminus o}})
\end{split}\\
    \text{with } \alpha=(1 - \sum_{o \in O} \frac{cf_o^{\Tau^{tgt}}}{pf_o^{\Tau^{ref}}})  \text{ and } \beta^l=\frac{cf_o^{\Tau^{tgt}}}{pf_o^{\Tau^{ref}}}
\end{multline}

where we find that:
\begin{equation}
\label{eqn:alphas}
    \alpha = 1 - \sum_{o \in O} \beta^o
\end{equation}
Substituting $\beta^o$ for $cf_o^{\Tau^{tgt}}$ in (\ref{eqn:linear_system_line_disconnection}), we have:

\begin{multline}
\label{eqn:LODF_system_solve}
    pf_l^{\Tau^{ref}} + \sum_{o \in O } lodf_{o,l} \times  \beta^o pf_o^{\Tau^{ref}} =0 \text{ for all } l \in O\\
    pf_l^{\Tau^{ref}} = - \sum_{o \in O } lodf_{o,l} \times  \beta^o pf_o^{\Tau^{ref}} \text{ for all } l \in O
\end{multline}

And reusing LODF definition from (\ref{eqn:LODF}) we recover (\ref{eqn:ST_linear_system}):
\begin{multline}
\label{eqn:sup_topo_theorem_linear_system_deco_betas}
     pf_l^{\Tau^{ref}} = \beta^{l}\times pf_l^{\Tau^{ref}} + \sum_{o \in O,o \neq l }(\frac{pf_{l}^{\Tau^{ref}} - pf_{l}^{\Tau^{\tau_{\setminus o}}}}{pf_{o}^{\Tau^{ref}}})\beta^{o} pf_o^{ref} \\
     1 = \beta^{l}\times 1 + \sum_{o \in O,o \neq l }(1 - \frac{ pf_{l}^{\Tau^{\tau_{\setminus o}}}}{pf_{l}^{\Tau^{ref}}})\beta^{o} \text{ for all } l \in O
     \\
     1 = \beta^{l}\times 1 + \sum_{o \in O,o \neq l }tsf_l^{\Tau^{\tau_{\setminus o}}}\beta^{o} \text{ for all } l \in O
\end{multline}

We see from (\ref{eqn:topo_theorem}) or (\ref{eqn:sup_topo_theorem}) that it solely relies here on reusing known quantities $PF(\Tau^{ref})$, $PF(\Tau^{\tau_{\setminus o}})$, and requires one to solve an additional linear system of equations of the size of the number of topological changes $n_{\tau}$, that is the number of line disconnections here $n_{O}$.

\section{EST for Line reconnections}
\label{sec:line_reco}

In this section, we consider line reconnections $\tau_{=l}$ for lines $l \in R$  to reach a target topology \textbf{$\Tau^{tgt}=\Tau_{R}$} departing from a reference topology \textbf{$\Tau^{ref}=\Tau_{O}$} in which those lines were initially not connected. 
So we change states in a reverse order from Figure \ref{fig:cancelling_flow}: the reference state is the one with disconnected lines at the bottom, and the target state is with connected lines as at the top. Here we cannot reuse the linear system (\ref{eqn:linear_system_line_disconnection}) as is, as there is no equivalent of $pf^{ref}_l$ for initially disconnected lines $l$ in this case. However, we can reuse the EST for line disconnections from previous section \ref{sec:line_disco_th}, to prove it for reconnections such that:
\begin{multline}
\label{eqn:sup_topo_theorem_reco_lines}
    PF(\Tau^{tgt}) = \alpha_R PF(\Tau^{ref}) + \sum_{l_i \in R} \beta_R^{l_i} PF(\Tau^{\tau_{=l_i}})\\
        \text{with } \alpha_R=1-\sum_{l_i \in R}\beta_R^{l_i}
\end{multline}

In the following, we will first demonstrate that this superposition exists and is unique. We will further identify the underlying linear system to solve, to derive the coefficients.

\subsubsection{Demonstration}

We first demonstrate the EST unicity and further its existence.

\renewcommand\qedsymbol{$\blacksquare$}
\begin{proof} \textbf{Unicity}.
From (\ref{eqn:sup_topo_theorem_reco_lines}), we can deduce that, when $\tau_{=l_i}$ has an effect resulting in non-null $pf_{l_i}^{\Tau^{\tau_{=l_i}}}$, the $\beta$ coefficient would have a unique value as simple as:
\begin{equation}
\label{eqn:sup_topo_theorem_reco_beta_value}
    \beta_R^{l_i}=\frac{pf_{l_i}^{_{\Tau^{tgt}}}}{pf_{l_i}^{\Tau^{\tau_{=l_i}}}}
\end{equation}
This is due to line $l_i$ flow being non-null only in topologies $\Tau^{tgt}$ and $\Tau^{\tau_{=l_i}}$ in which the line is connected. For all the other topologies, $l_i$ is disconnected with null flow. Hence all other terms for $l_i$ index are null in EST for line reconnections. 

Given the uniqueness of powerflows, this proves the uniqueness of $\beta$ coefficients, as well as $\alpha$ by transitivity. Note however that our objective is to compute $pf_{l_i}^{_{\Tau^{tgt}}}$ after deriving the coefficients $\beta_R^{l_j}$. We should find later a coefficients derivation based on the reference state instead of the target state.

When $pf_{l_i}^{\Tau^{\tau_{=l_i}}}$ is null, this is the case when reconnecting this single line does not affect the reference state, resulting in the same reference state $PF(\Tau^{ref})$. Hence $\alpha_R$ and $\beta_R^{l_i}$ can be merged, such as setting $\beta_R^{l_i}=0$.
\end{proof}

\begin{proof} \textbf{Existence.}
We first consider the reconnection of two lines $\{l_1,l_2\}$ from a topology $\Tau_O$ in which both are disconnected to reach a topology $\Tau_R$ in which both are reconnected. 

We can start by reusing (\ref{eqn:sup_topo_theorem}) with reverse roles of reference and target topologies:
\begin{multline}
\label{eqn:sup_topo_theorem_reco_2_lines_start}
    PF(\Tau_{O}) = \alpha_{O} \times PF(\Tau_{R}) +  \sum_{i \in \{1,2\}} \beta^{l_i}_{O} \times PF(\Tau^{\tau_{\setminus l_i}}_{R})\\
    PF(\Tau_{O}) = \alpha_{O} \times PF(\Tau_{R}) +  \beta^{l_1}_{O} \times PF(\Tau^{\tau_{=l_2}}_{R}) +\beta^{l_2}_{O} \times PF(\Tau^{\tau_{= l_1}}_{R})
\end{multline}

Indeed, in the 2 lines reconnection case, the unitary reconnection grid topology $\Tau^{\tau_{=l_1}}_{O}$ under line $l_1$ reconnection $\tau_{=l_1}$, is the same as the unitary disconnection grid topology $\Tau^{\tau_{\setminus l_2}}_{R}$ under line disconnection $l_2$, since $\Tau_{R}\circ \tau_{ \setminus l_2} = \Tau_{O}\circ \tau_{=l_1}$. 

So $PF(\Tau^{\tau_{\setminus l_2}}_{R}) = PF(\Tau^{\tau_{=l_1}}_{O})$ and $PF(\Tau^{\tau_{\setminus l_1}}_{R}) = PF(\Tau^{\tau_{=l_2}}_{O})$.

This time $PF(\Tau_{R})=PF(\Tau^{tgt})$ is the superposed state we are looking for. So rearranging (\ref{eqn:sup_topo_theorem_reco_2_lines_start}) leads to:

\begin{multline}
\label{eqn:sup_topo_theorem_reco_2_lines}
    PF(\Tau_{R}) = \frac{1}{\alpha_{O}}(PF(\Tau_{O}) - \beta^{l_1}_{O} \times PF(\Tau^{\tau_{=l_2}}_{R}) - \beta^{l_2}_{O} \times PF(\Tau^{\tau_{= l_1}}_{R})) \\ 
    PF(\Tau_{R}) = \alpha_{R} \times PF(\Tau_{O}) + \sum_{i \in \{1,2\}}\beta_{R}^{l_i} \times PF(\Tau^{\tau_{=l_i}}_{O}) \\
    PF(\Tau^{tgt}) = \alpha_{R} \times PF(\Tau^{ref}) + \sum_{i \in \{1,2\}}\beta_{R}^{l_i} \times PF(\Tau^{\tau_{=l_i}}) \\
    \text{with } \alpha_{R}=\frac{1}{\alpha_{O}} \text{, } \beta_{R}^{l_1}=\frac{-\beta^{l_2}_{O}}{\alpha_{O}} \text{, } \beta_{R}^{l_2}=\frac{-\beta^{l_1}_{O}}{\alpha_{O}} 
\end{multline}

From (\ref{eqn:alphas}), we have:
\begin{equation}
    \frac{1}{\alpha_{O}} = 1 + \sum_{i \in \{1,2\}} \frac{-\beta^{l_i}_{O}}{\alpha_{O}}
\end{equation}

So we recover:
\begin{equation}
\label{eqn:alphas_2_lines_reco}
    \alpha_{R} = 1 - \sum_{i \in \{1,2\}} \beta^{l_i}_{R}
\end{equation}

This works out when $\alpha_{O}$ is non null. As $\alpha_R$ and $\beta^{l_i}$ are all properly defined from (\ref{eqn:sup_topo_theorem_reco_beta_value}) and cannot be infinite, $\alpha_{O}$ is indeed never null.  
For more than two line reconnections, we can \textbf{recursively} apply (\ref{eqn:sup_topo_theorem_reco_2_lines}). We leave this complimentary demonstration to the Appendix \ref{appendix:multi_reco}. By applying the same recursion, we find for any number of line reconnection the existence of an EST:

\begin{equation}
\label{eqn:sup_topo_theorem_all_reco}
PF(\Tau^{tgt}) = \alpha_{R} PF(\Tau^{ref}) + \sum_{l_i \in R} \beta_{R}^{l_i} PF(\Tau^{\tau_{=l_i}})
\end{equation}

\end{proof}
\vspace{0.2cm}

\subsubsection{Linear system of equations to solve}
\label{sec:eqn_system_reconnection}
As mentioned before, we need to derive the coefficients from the reference state in which $pf^{ref}_l$ are undefined for initially disconnected lines $l \in R$. 
We will use the EST we just demonstrated in (\ref{eqn:sup_topo_theorem_all_reco}) to identify more specifically the linear system of equations to solve in this case. It will be based on voltage angle difference at line extremities of interest. The quantity we indeed know in the reference state and unitary reconnection states are virtual induced flows such as $vf_l^{\Tau _O} =  \sigma_l \Delta \theta_l^{\Tau_O}$. Our objective is to derive equations based on these terms.

Let's hence consider once again the canceling flow equivalent model of working with the fully connected topology all along in which from (\ref{eqn:canceling_flow_equation}) we have: 
\begin{equation}
\label{eqn:null_flow_decomposition}
pf_l^{\Tau _O}=vf_l^{\Tau _O} - cf_l^{\Tau _O} = 0
\end{equation}



Given that $pf_l^{\Tau _O}$ and $pf_l^{\Tau_O^{\tau_{=l_j}}}$ for $l_i \in R_{\setminus l}$ are null, we have $n_R=card(R)$ equations for each line $l \in R$ to reconnect:
\begin{equation}   
\label{eqn:reconnect_raw_equation}
    \alpha_R pf^{\Tau_O}_{l} + \sum_{l_i \in R_{\setminus l} } \beta_R^{l_i} pf^{\Tau _O^{\tau_{=l_i}}}_{l} = 0
\end{equation}

To make $vf_l$ terms appear, we substitute (\ref{eqn:null_flow_decomposition}) leading to:
\begin{equation}
\label{eqn:sup_th_reco_injected_flow_Rancelling_eq}
     \alpha_R (vf^{\Tau_O}_{l} - cf^{\Tau_O}_{l}) + \sum_{l_i \in R_{\setminus l} }\beta_R^{l_i} (vf^{\Tau _O^{\tau_{=l_i}}}_{l} - cf^{\Tau _O^{\tau_{=l_i}}}_{l}) = 0
\end{equation}

Rearranging it as a sum of canceling flows and a sum of virtual flows, we get:
\begin{multline} 
\label{eqn:sup_th_reco_injected_flow_Rancelling_eq}
     ( \alpha_R vf^{\Tau_O}_{l}+\sum_{l_i \in R_{\setminus l} }\beta_R^{l_i}  vf^{\Tau _O^{\tau_{=l_i}}}_{l}) - \\
     (\alpha_R cf^{\Tau_O}_{l} + \sum_{l_i \in R_{\setminus l}}\beta_R^{l_i} cf^{\Tau _O^{\tau_{=l_i}}}_{l}) = 0
\end{multline}

We leave to the Appendix \ref{appendix:cancel_out} the complementary demonstration that injected canceling flows, at each line to reconnect, cancel out through the superposition of grid states in ST. The second term being null, the first part is also null. From (\ref{eqn:sup_th_reco_injected_flow_Rancelling_eq}) we are left with our linear system of equations with only known $vf_l$ terms:

\begin{multline}
\begin{split}
\alpha_R vf^{\Tau_O}_{l}+\sum_{l_i R_{\setminus l} }\beta_R^{l_i}  vf^{T_O^{\tau_{=l_i}}}_{l} & = 0 \\
\alpha_R vf^{\Tau^{ref}}_{l}+\sum_{l_i \in R_{\setminus l} }\beta_R^{l_i}  vf^{T^{\tau_{=l_i}}}_{l} & = 0 \text{ with $\Tau^{ref}=T_O$}\\
\end{split}
\end{multline}

By substituting (\ref{eqn:virtual_induced_flow}) we have:
\begin{multline}
\begin{split}
\label{eqn:sup_th_reco_induced_flow_Rancelling_eq}
\alpha_R \sigma_l\Delta \theta^{\Tau^{ref}}_{l}+\sum_{l_i \in R_{\setminus l} }\beta_R^{l_i}  \sigma_l\Delta \theta^{T_O^{\tau_{=l_i}}}_{l} & = 0 \\
\alpha_R \Delta \theta^{\Tau_O}_{l}+\sum_{l_i \in R_{\setminus l} }\beta_R^{l_i}  \Delta \theta^{T_O^{\tau_{=l_i}}}_{l} & = 0 \\
\alpha_R+\sum_{l_i \in R_{\setminus l} }\beta_R^{l_i} \frac{\Delta \theta^{T_O^{\tau_{=l_i}}}_{l}}{\Delta \theta^{\Tau_O}_{l}} & = 0 
\end{split}
\end{multline}

if $\Delta \theta^{\Tau_O}_{l}$ is non-null. In the null case, line $l$ can be considered reconnected with null flow $pf_{l}^{\Tau^{\tau_{=l}}}=pf^{\Tau_O}_{l}=0$ and $\beta_R^l=0$ as in (\ref{eqn:sup_topo_theorem_reco_beta_value}), hence discarding it from the system of equations. 

By reusing (\ref{eqn:alphas_3_lines_reco}), we have for all $l \in R$:
\begin{equation}
     (1 - \sum_{l_i \in R} \beta^{l_i}_{R}) + \sum_{l_i \in R_{\setminus l} }\beta_R^{l_i} \frac{\Delta \theta^{T_O^{\tau_{=l_i}}}_{l}}{\Delta\theta^{\Tau_O}_{l}} = 0
\end{equation}

\begin{multline}
\label{eqn:sup_th_reco_delta_theta_Rancelling_eq_simplyfied}
\begin{split}
     \beta^{l}_{R}\times 1 + \sum_{l_i \in R_{\setminus l} }\beta_R^{l_i}(1 - \frac{\Delta \theta^{T_O^{\tau_{=l_i}}}_{l}}{\Delta\theta^{\Tau_O}_{l}}) = 1\\
    \beta^{l}_{R}\times 1 + \sum_{l_i \in R_{\setminus l} }tsf_l^{T_O^{\tau_{=l_i}}}\beta_R^{l_i} = 1
\end{split}
\end{multline}

%

where we recover a system of equations in the form of (\ref{eqn:ST_linear_system}).

\subsubsection{Interpretation - canceling voltage angle difference equivalent model and LCDF }

We find the same equations as for line disconnections, but considering $\Delta\theta_l$ instead of $pf_l$.
An equivalent model is hence of canceling voltage angle difference instead of canceling flows. We recover the physical reasoning of a line reconnection equivalent model from \cite{Sauer2001}:
\begin{itemize}
    \item starting with the line disconnected in reference topology $\Tau_O$, inject some nodal power that cancels out the voltage angle difference $\Delta \theta_l^{\Tau_O}$ at line $l$ when disconnected originally
    \item then physically reconnect line $l$. This does not induce any effect on the state as its flow remains null given that the voltage angle difference has been canceled out just before. It is closed without impacting the system.
    \item finally remove the injected balance power to retrieve the desired target state, that is with line $l$ reconnected
\end{itemize}

The problem hence reduces to finding those voltage angle difference injections. The same equations as for line disconnections therefore exist, one solely needs to replace $pf_l$ by $\Delta\theta_l$. So analogously to LODF, we can define LCDF (line closing distribution factors) such as:

\begin{equation}
    LCDF_{r,l} = \frac{\Delta\theta_l^{\Tau^{\tau_{=r}}}- \Delta\theta_l^{\Tau^{ref}}}{\Delta\theta_r^{\Tau^{ref}}}
\end{equation}

And similarly to (\ref{eqn:LODF_system_solve}) the system of equations to solve could be rewritten and interpreted as:

\begin{equation}
    \Delta\theta^{\Tau_O}_{l} + \sum_{r \in R} LCDF_{r,l} \times \beta^r\Delta\theta^{\Tau_O}_{r} = 0 
\end{equation}


%
%
%

%
%
%
\section{Power flow calculation with combined changes}
\label{sec:power_flow_Rombined}
\subsection{Combined line disconnections and reconnections}
\label{sec:line_disco_reco_combined}

%

We finally demonstrate that combined line disconnections and reconnections fall under a single linear system to solve.
At first sight, the two kinds of linear systems seem indeed compatible as they both fall under the same linear system form (\ref{eqn:ST_linear_system}) as from equations (\ref{eqn:sup_topo_theorem_linear_system_deco_betas}) and (\ref{eqn:sup_th_reco_delta_theta_Rancelling_eq_simplyfied}).

\subsubsection{Demonstration}
Let us first deal with the single combination case and generalize it to the multi-combinatorial case.

\begin{proof}\textbf{Existence and Unicity.}
As for the two-line reconnection case, in the combined one-line disconnection one-line reconnection case we can start by reusing (\ref{eqn:sup_topo_theorem}) with reverse roles: reference and target topologies become unitary change topologies and conversely. Here we start with $l_1$ connected that we disconnect and $l_2$ disconnected that we reconnect. So the topology with both lines disconnected is $\Tau^{\tau_{\setminus l_1}}$ and the topology with both lines connected is $\Tau^{\tau_{= l_2}}$ such that:

\begin{equation}
\label{eqn:sup_topo_theorem_1_reco_1_disco_init}
    PF(\Tau^{\tau_{\setminus l_1}}) = \alpha_{O}PF(\Tau^{\tau_{=l_2}})+  \beta^{l_1}_{O} PF(\Tau^{tgt}) +  \beta^{l_2}_{O} PF(\Tau^{ref})
\end{equation}

The topology corresponding to $\beta^{l_1}_{O}$, $\Tau^{tgt}$, is the one for which $l_1$ only is disconnected, and similarly for $\beta^{l_2}_{O}$.
By rearranging equation (\ref{eqn:sup_topo_theorem_1_reco_1_disco_init}), we retrieve the EST theorem:

\begin{multline}
\label{eqn:sup_topo_theorem_1_reco_1_disco_final}
\begin{split}
    PF(\Tau^{tgt}) & = \frac{1}{\beta^{l_1}_{O}}PF(\Tau^{\tau_{\setminus l_1}})-\frac{\alpha_{O}}{\beta^{l_1}_{O}}PF(\Tau^{\tau_{=l_2}})-\frac{\beta^{l_2}_{O}}{\beta^{l_1}_{O}}  PF(\Tau^{ref}) \\
    & = \alpha PF(\Tau^{ref}) + \beta^{l_1}PF(\Tau^{\tau_{\setminus l_1}}) + \beta^{l_2}PF(\Tau^{\tau_{=l_2}})
\end{split}\\
    \text{with } \alpha=\frac{-\beta^{l_2}_{O}}{\beta^{l_1}_{O}} \text{ , } \beta^{l_1}=\frac{1}{\beta^{l_1}_{O}} \text{ , } \beta^{l_2}=\frac{-\alpha_{O}}{\beta^{l_1}_{O}}
\end{multline}

As $\beta^{l_2}_{O}=1-\beta^{l_1}_{O}-\alpha_{O}$ from (\ref{eqn:alphas}), we recover:
\begin{multline}
\begin{split}
    \alpha & =\frac{(\alpha_{O}+\beta^{l_1}_{O}-1)}{\beta^{l_1}_{O}}
    & = \frac{\alpha_{O}}{\beta^{l_1}_{O}} +1 - \frac{1}{\beta^{l_1}_{O}} \\
    & = 1 - \beta^{l_1} - \beta^{l_2}
\end{split}
\end{multline}
The unicity of $\alpha$ and $\beta^{l_i}$ stems from the unicity of $\alpha_O$ and $\beta^{l_i}_O$ previously proven.

For the multi-combinatorial case, one can use
successive recursion using EST for line disconnections (\ref{eqn:sup_topo_theorem}), reconnections (\ref{eqn:sup_topo_theorem_reco_lines}) and one reconnection - one disconnection combined (\ref{eqn:sup_topo_theorem_1_reco_1_disco_final}), EST is derived similarly to the demonstration for multiple-line reconnections as in Appendix \ref{appendix:multi_reco}:

\begin{multline}
\label{eqn:sup_topo_theorem_Rombined_reco_deco}
    PF(\Tau^{tgt}) = \alpha PF(\Tau^{ref}) + \sum_{l_i \in R} \beta_R^{l_i} PF(\Tau^{\tau_{=l_i}})\\
    +\sum_{l_j \in O} \beta_O^{l_j} PF(\Tau^{\tau_{\setminus l_j}})\\
        \text{with } \alpha=1-\sum_{l_i \in R}\beta_R^{l_i}-\sum_{l_j \in O}\beta_O^{l_j}
\end{multline}
\end{proof}

\subsubsection{Linear System to solve}
We reuse (\ref{eqn:sup_topo_theorem_Rombined_reco_deco}), first along line reconnection indices, and then along line disconnection indices, to derive our set of equations.

As in (\ref{eqn:reconnect_raw_equation}) for line reconnections only, we have here:
\begin{multline}
\begin{split}
    \text{for all  l $\in$ R, } pf^{\Tau^{tgt}}_{l}  - \beta_R^{l}pf^{\Tau^{\tau_{=l}}}_{l} & = 0  \\
    \alpha pf^{\Tau^{ref}}_{l} + \sum_{l_i \in R_{\setminus l} } \beta_R^{l_i} pf^{\Tau^{\tau_{=l_i}}}_{l} + \sum_{l_j \in O} \beta_O^{l_j} pf^{\Tau^{\tau_{\setminus l_j}}}_{l} & = 0
\end{split}
\end{multline} 
In the second equation, all power flows for line $l$ are null as the line is disconnected in all those states. By reapplying the canceling flow equivalent model reasoning from equations (\ref{eqn:sup_th_reco_injected_flow_Rancelling_eq}) up to (\ref{eqn:sup_th_reco_delta_theta_Rancelling_eq_simplyfied}) we derive a first set of $n_R$ independent equations:

\begin{equation}
\label{eqn:sup_th_reco__deco_Rombined_delta_theta_reco}
     \beta^{l}_{R}\times 1 + \sum_{l_i \in R_{\setminus l} }\beta_R^{l_i}(1 - \frac{\Delta \theta^{\Tau^{\tau_{=l_i}}}_{l}}{\Delta\theta^{\Tau^{ref}}_{l}})+\sum_{l_j \in O}\beta_O^{l_j}(1 - \frac{\Delta \theta^{\Tau^{\tau_{\setminus l_j}}}_{l}}{\Delta\theta^{\Tau^{ref}}_{l}}) = 1
\end{equation}

\begin{equation}
\label{eqn:sup_th_reco__deco_Rombined_delta_theta_reco}
     \beta^{l}_{R}\times 1 + \sum_{l_i \in R_{\setminus l} }\beta_R^{l_i}tsf_{l}^{\Tau^{\tau_{=l_i}}}+\sum_{l_j \in O}\beta_O^{l_j}tsf_{l}^{\Tau^{\tau_{\setminus l_j}}} = 1
\end{equation}

For equations related to each line disconnections we have:

\begin{multline}
\begin{split}
    \text{for all  l $\in$ O, } pf^{\Tau^{tgt}}_{l}  & = 0  \\
    \alpha pf^{\Tau^{ref}}_{l} + \sum_{l_i \in R} \beta_R^{l_i} pf^{\Tau^{\tau_{=l_i}}}_{l} + \sum_{l_j \in O} \beta_O^{l_j} pf^{\Tau^{\tau_{\setminus l_j}}}_{l} & = 0 \\ 
    \alpha pf^{\Tau^{ref}}_{l} + \sum_{l_i \in R} \beta_R^{l_i} pf^{\Tau^{\tau_{=l_i}}}_{l} + \sum_{l_j \in O,l_j \neq l } \beta_O^{l_j} pf^{\Tau^{\tau_{\setminus l_j}}}_{l} & = 0
\end{split}
\end{multline}
The last equation stems from the fact that $pf^{\Tau^{\tau_{\setminus l}}}_{l}=0$.
Those power flow quantities are known as the line is connected in these states. We hence derive a last set of $n_O$ independent equations by replacing $\alpha$ with $\beta$s:
\begin{equation}
\label{eqn:sup_th_reco__deco_Rombined_delta_theta_deco}
     \beta^{l}_{O}\times 1 + \sum_{l_i \in R}\beta_R^{l_i}(1 - \frac{pf^{\Tau^{\tau_{=l_i}}}_{l}}{pf^{\Tau^{ref}}_{l}})+\sum_{l_j \in O, l_j \neq l }\beta_O^{l_j}(1 - \frac{pf^{\Tau^{\tau_{\setminus l_j}}}_{l}}{pf^{\Tau^{ref}}_{l}}) = 1
\end{equation}

\begin{equation}
     \beta^{l}_{O}\times 1 + \sum_{l_i \in R}\beta_R^{l_i}tsf_{l}^{\Tau^{\tau_{=l_i}}}+\sum_{l_j \in O, l_j \neq l }\beta_O^{l_j}tsf_{l}^{\Tau^{\tau_{\setminus l_j}}} = 1
\end{equation}

Those two sets of equation can eventually be merged under a single system to solve:
\begin{equation}
\label{eqn:line_reco_disco_system}
     \beta^{l}\times 1 + \sum_{l_i, l_i \neq l}\beta^{l_i}tsf_{l}^{\Tau^{\tau_i}} = 1
\end{equation}

Note that when $pf_l^{ref}$ is non-null, $\frac{pf^{\Tau^{\tau_j}}_{l}}{pf^{\Tau^{ref}}_{l}}$ can alternatively be used instead of $\frac{\Delta \theta^{\tau_j}_{l}}{\Delta\theta^{\Tau^{ref}}_{l}}$ and conversely when $\Delta\theta^{\Tau^{ref}}_{l}$ is non-null. The linear system to solve is again of the form of (\ref{eqn:ST_linear_system}).

\subsection{Node splitting and merging topological changes}
A node-splitting change can be modeled through a null-impedance virtual line disconnection \cite{goldis2016shift} in between the two target nodes and conversely a node merging as a virtual line reconnection. Physically, one could represent this virtual line as a coupling breaker opened or closed between 2 bus bars that represents the nodes. Previous ST demonstrations and systems of linear equations directly apply to those changes as no hypothesis or usage of grid properties such as line impedance were made, with only reliance on grid state knowledge. 

For node merging, $\Delta \theta_{nodes}$ between the two nodes to be merged can be used in the equations. For node splitting, the flow $pf_{nodes}^{T}$ through the null-impedance line virtually connecting the two virtual nodes, not yet split, needs to be computed. It can be done based on line flows at the substation which results in a residual power flow at each virtual node, as they are not split yet:
\begin{equation}
    pf_{nodes}^{T}= \sum_{l_i \in node_1}pf_{l_i}^{T} 
\end{equation} 

As node splitting can be assimilated to line disconnection and node merging to line reconnection, they fall under the same system (\ref{eqn:line_reco_disco_system}) which hence integrates all four types of topological changes. 

\section{Experiments \& Application Results}
\label{sec:results}

\antoine{expé en plus: pf action unitaire en AC, combinaison via theorem sup ? Une baseline ?

Aussi voir comment se compare notre méthode à GLODF classique ?}
In this section, we validate the accuracy and effectiveness of EST implementation and prove its practical interest for speeding up two first applications.
The source code is open-source on Github\footnote{https://github.com/marota/Topology\_Superposition\_Theorem.git}. It uses Grid2Op library \cite{grid2op} to apply topological changes and LightSim2Grid \cite{lightsim2grid} as a reference power flow solver that offers speeds comparable to industrial ones. 

\subsection{EST numerical implementation and utility}
\subsubsection{EST Numerical validation}

\begin{table}[]
    \centering
    \begin{tabular}{|c|c|c|c|c|c|c|c|c|c|}
         \hline 
          & \multicolumn{8}{|c|}{EST coefficients for topological changes $\tau$} & speed\\
         $n°$ & \multicolumn{2}{|c|}{$l_{1-3}$} & \multicolumn{2}{|c|}{$l_{9-10}$}  & \multicolumn{2}{|c|}{$sub_4$}  & \multicolumn{2}{|c|}{$sub_5$} & -up 
         \\

          & O & R & O & R & S & M & S & M & \\
         \hline
         0 & 1.02 &  & 1.09 &  &  &  &  & & 5.0 \\\hline 
         1 & & 0.98 &  & 0.92 &  &  &  &  & 3.2  \\\hline 
         2 & & &  & & 0.62 &  & 0.92 &  & 4.0  \\\hline 
         3 & & &  & &  & 1.69 &  & 1.14 & 6.6 \\\hline 
         5 &  & 1.19 & 0.45 & & 0.69 &  & & 1.59 & 2.2  \\\hline 
    \end{tabular}
    \caption{On IEEE14, the EST coefficients and speedup factors compared to power-flow are computed for 5 different combined action configurations. There are 4 different combined actions of same topological change type (line outage (O) or reconnection (R), bus splitting (S) or merging (M)) and a slower all type combined action. $l_{1-3}$ is the line connecting substations 1 to 3, and $sub_4$ is substation n°4.} 
    \label{tab:combined_Rhanges}
\end{table}

To evaluate the equivalence of the EST method compared to standard powerflow resolution, we select the combined actions of disconnecting and connecting lines, and splitting and merging buses, for the simple IEEE 14 grid as shown in Table \ref{tab:combined_Rhanges}. 
Configuration n°3 is the same as for Figure  \ref{fig:node_spliiting_example} with combined bus splittings at substations 4 and 5. For a combination of the same action type or different action types, we solve the linear systems of EST and find the displayed $\beta$ coefficients. Using EST, we further confirm the same flow values as usual DC power flows in all those cases with at least 4 decimals. 

\subsubsection{Interpretability of Combined Actions}
Given the complexity of interactions between power grid structures, it is important to understand the behavior of a topological action. Such understanding makes the selection of corrective or preventive action easier, which can facilitate and accelerate the operator's decision-making. To understand how the EST helps in this problem, we select two disconnection use cases: 
\begin{itemize}
    \item Two lines from separate clicks ($l_{1-2}$ and $l_{11-12}$), which are electrically distant from each other.
    \item Two lines from the same click($l_{1-2}$ and $l_{1-3}$), which are electrically near to each other.
\end{itemize}

As mentioned previously, the power flow through the remaining transmission lines can be calculated using the $\beta$s coefficients. We have calculated the values of $\beta$ for each case in Table \ref{tab:beta_values}. Note that when the disconnected lines come from different clicks, $\beta$s is close to unity ($\beta \simeq 1$), which means that flow redistribution to any remaining power line amounts to disconnecting each line independently from the other. Consequently, when the values of $\beta$ are identical to the identity, the actions performed are electrically distant and can be considered as independent actions.

In the second case, where the actions are applied within the same click, the corresponding $\beta$ values deviate from the identity. This is due to the proximity of the topological changes interacting with one another.

Therefore, the interpretation of $\beta$s can help clarify the independence of the applied action or the predominance of some actions. The same interpretation can be used for line reconnections, bus splits or merges, and even for mixed actions between all these elements.


\begin{table}[]
    \centering
    \begin{tabular}{|c|c|c|}
         \hline
         combined line outages&$\beta_1$& $\beta_2$ \\
         \hline same click lines
         $l_{1-2}$, $l_{1-3}$ &1.02&1.00\\
         \hline separate click lines
         $l_{1-2}$, $l_{11-12}$&1.52&1.63\\
         \hline
    \end{tabular}
    \caption{For combined line disconnections, the $\beta$s EST coefficients  close to 1 in separate clicks and far from 1 in same clicks because of mutual influence.}
    \label{tab:beta_values}
\end{table}

\subsubsection{Improved AC load-flow approximation}

\begin{figure}[htbp]
    \centering
    \includegraphics[width=0.45\textwidth]{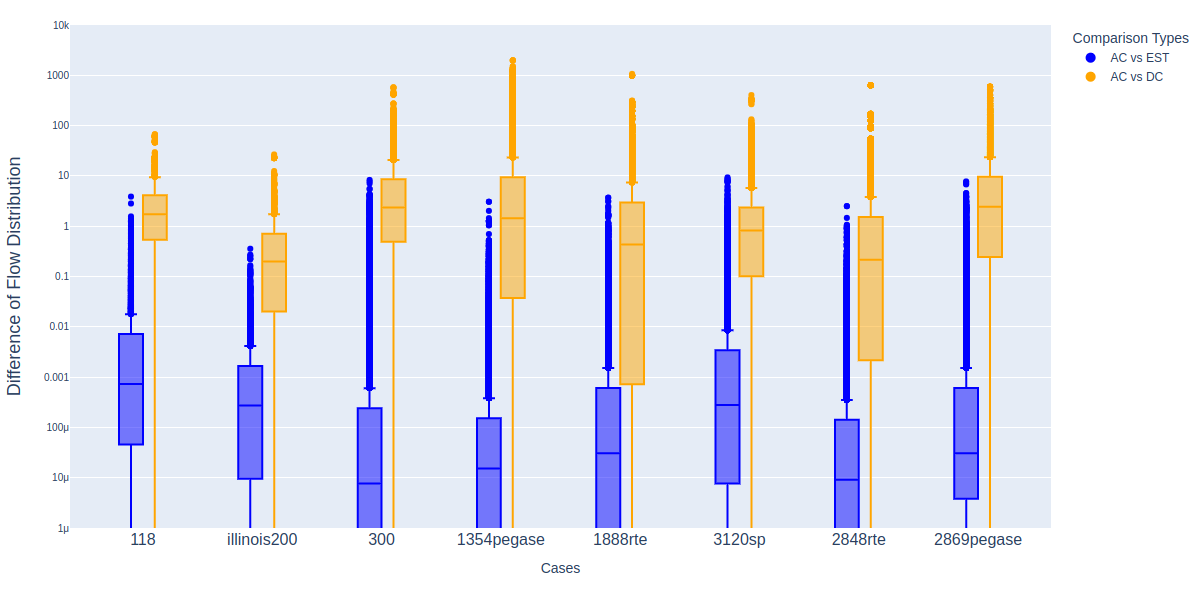}
    \caption{The box plots of accuracy error distributions to AC load flow for DC approximation (orange) and the EST method (blue) in the case of influenced actions.}
    \label{fig:flow_diff}
\end{figure}

This study evaluates the accuracy of the EST model when applied as is to active AC power flows. Specifically, we analyze power flow results for a combinatorial depth of 3 line disconnections, comparing line flows obtained from AC computation, the DC approximation, and the EST model using AC pre-computed reference and unitary single-line disconnection states. We distinguish two cases: the simpler independent actions case, where $\beta\approx1$ for all $\beta$s, and the more challenging influenced actions case, where $|\beta-1|>=0.1$ for all $\beta$s. For each grid case, we assess $100$ different combinations of dependent and independent actions. For the independent actions case, the maximum error of line flow between EST and AC remains of the order of $0.1MW$ in all cases. Figure \ref{fig:flow_diff} presents the results for the influenced actions. The difference between EST and AC flows remains almost always below the acceptable threshold of $1MW$ in all cases, except for very few points up to $10MW$ difference. In contrast, the standard DC approximation has an interquartile error range that often goes beyond $10MW$ with extreme values of about $100MW$ difference. 
In this challenging case with possibly large non-linear effects between combined actions, the EST appears to be a good and robust approximation of AC powerflow. Future work could study the larger differences possibly due to greater voltage and connectivity drops, and investigate the overall generalization of EST to AC power flows.

\subsection{Two applications illustrated}
\subsubsection{Remedial Action search}
When looking for remedial actions, one topological change might not be enough, and one might need to combine a few of them. Topological change in Table \ref{tab:combined_Rhanges} could be possible remedial actions. We hence see on a small grid the speed-up factors we can obtain while computing their combinatorial effect compared to state-of-the-art power flow LightSim2Grid \cite{lightsim2grid}. The speed-up however decreases while the number of unitary topological changes increases as we can see for configuration n°5. For larger grids of more than a thousand nodes, we see from Table \ref{tab:number_actions} that more than an order of magnitude speed-up can be obtained for 2 actions, or even a few more, combined.  

To assess its relevance for larger grids, we estimate the action combinatorial depths until EST remains faster than this DC power flow. Table \ref{tab:number_actions}  also highlights the maximum number of actions for each grid case. A significantly high combinatorial depth is observed for large grids, often exceeding $25$ combined actions. This is quite substantial in an operational context as the number of allowed manoeuvers for an overall remedial action does not exceed 10 for instance today. This shows that EST should always be preferred to standard DC powerflow for this application.

\begin{table}[]
    \centering
    \scalebox{0.8}{
    \begin{tabular}{|c|c|c|c|c|c|c|c|}
        \hline
        Grid Case&118&300&1354peg&1888rte&3120sp&2848rte& 2869peg\\
        \hline
         Speed-up factor&3.1&6.2&24.4&28.1&47.6&42.7&51.0\\
         \hline
         Action Depth&10&16&22&26&27&29&29\\

         \hline
    \end{tabular}}
    \caption{Over 100 action trials for each grid case, the EST median speed-up factor for 2 actions combined, and median number of the combinatorial action depth until which EST remains faster than the power flow solver. }
    \label{tab:number_actions}
\end{table}

\subsubsection{Topological Action Security Analysis}

Security analysis (SA) is an application at the core of power system operations. It has been quite optimized over the years, reusing for instance the same KLU matrix factorization for all N-1 contingency computations. However, when a topological change is applied, one might need to be able to assess its robustness and recompute the SA quickly.

\begin{figure}[h]
    \centering
    \includegraphics[scale=0.24]{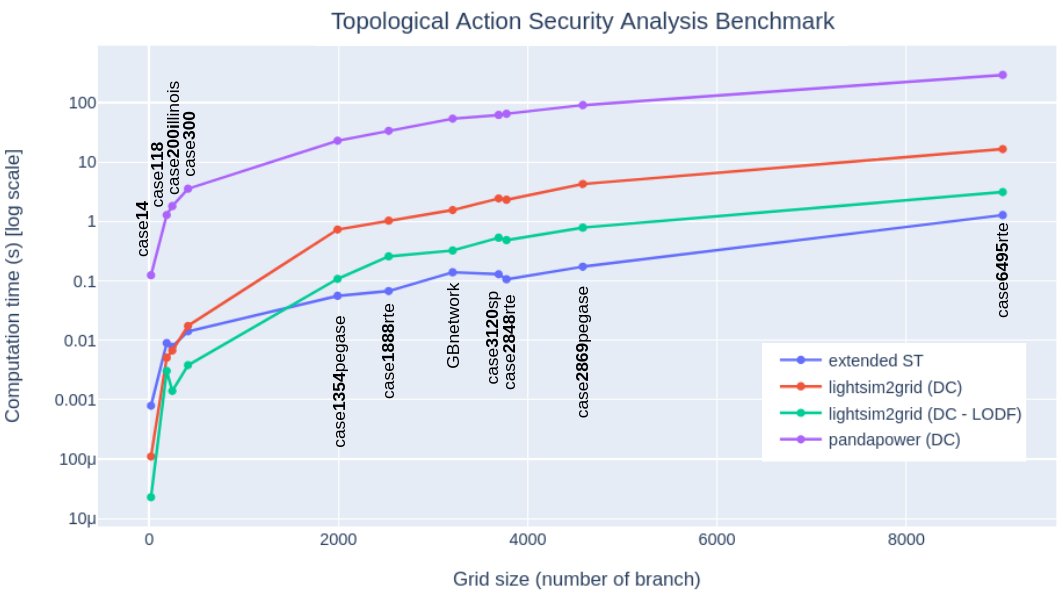}
    \caption{Comparing security analysis computation time in second (s) across grid sizes for EST and 2 reference DC power flow solvers LightSim2Grid \cite{lightsim2grid}, PandaPower \cite{thurner2018pandapower}. LODF is the most optimized computation specifically for DC.}
    \label{fig:security_analysis_Romputation_time}
\end{figure}

The EST can help reduce the computation time required for SA joined with topological actions. 
For experiments, for each grid under test, we select two random topological actions that do not break the grid. We then calculate the line outage SA for these two actions. Figure \ref{fig:security_analysis_Romputation_time} compares the computation time for the resulting SA. LightSim compiled C++ solver, once the topological change is computed and applied once, reuses the corresponding matrix factorization to optimize computation. This is hence a challenging baseline for our straightforward python implementation.

Our EST's python implementation is always several order of magnitude faster than pandapower, the python powerflow solver. For grids larger than 100 buses, the proposed approach exhibits faster computation compared to compiled power-flow solvers and scales linearly in the number of contingencies after this stage. We can observe that when considering a similar number of topological changes (a contingency plus topological action), the proposed technique solves an identical equation system regardless of the grid's size, explaining this scalability. It becomes at least an order of magnitude faster for grids larger than 1000 buses in its standard industrial SA version with KLU. Even for the most optimized C++ SA version that relies on PTDFs in the specific case of DC approximation, we still observe a speed-up factor of 5 or more. 

The computation time for SA can be improved by parallelization. This is achievable due to the independence of equation systems that need to be solved for each line outage simulation. Furthermore, leveraging the interpretability of $\beta$ coefficients could unlock further SA speed-ups by discarding more distant line disconnections that are detected as not influenced. Future work could study this more advanced use of EST.

\section{Conclusion}
\antoine{mention that one does not need to actually know the graph or complete topology, just individual changes}
In this paper, we have demonstrated the existence and unicity of an Extended Superposition Theorem for all types of unitary topological changes and their mixture. We have seen the speed-up and interpretability it can bring to power flow computations and their analysis. In particular, applications for remedial action search and security analysis can already benefit from it. We believe that it can be of very generic use and a foundation for improvements in many applications as well as integrated in optimization formulations, such as in optimal power flow. Future work will aim at revisiting some recent applications that involve topological changes in light of EST, such as grid segmentation \cite{Marot2018Guided, HENKA2022Segmentation}, leap net proxy power flow \cite{donon2020leap} or topological expert system \cite{marot2018expert}. It could lead to a better interpretation of models or results and possibly help improve the respective implementations. 

\bibliographystyle{ieeetr}
\bibliography{full_est_paper}

\begin{thebibliography}{10}

\bibitem{marot2022perspectives}
A.~Marot, A.~Kelly, M.~Naglic, V.~Barbesant, J.~Cremer, A.~Stefanov, and J.~Viebahn, ``Perspectives on future power system control centers for energy transition,'' {\em Journal of Modern Power Systems and Clean Energy}, vol.~10, no.~2, pp.~328--344, 2022.

\bibitem{little2021optimal}
E.~Little, S.~Bortolotti, J.-Y. Bourmaud, E.~Karangelos, and Y.~Perez, ``Optimal transmission topology for facilitating the growth of renewable power generation,'' in {\em 2021 IEEE Madrid PowerTech}, pp.~1--6, IEEE, 2021.

\bibitem{numan2020mobilizing}
M.~Numan, D.~Feng, F.~Abbas, S.~Habib, and A.~Rasool, ``Mobilizing grid flexibility through optimal transmission switching for power systems with large-scale renewable integration,'' {\em International Transactions on Electrical Energy Systems}, vol.~30, no.~3, p.~e12211, 2020.

\bibitem{marot2021learning}
A.~Marot, B.~Donnot, G.~Dulac-Arnold, A.~Kelly, A.~O’Sullivan, J.~Viebahn, M.~Awad, I.~Guyon, P.~Panciatici, and C.~Romero, ``Learning to run a power network challenge: a retrospective analysis,'' in {\em NeurIPS 2020 Competition and Demonstration Track}, pp.~112--132, PMLR, 2021.

\bibitem{fuxjager2023reinforcement}
A.~R. Fuxj{\"a}ger, K.~Kozak, M.~Dorfer, P.~M. Blies, and M.~Wasserer, ``Reinforcement learning based power grid day-ahead planning and ai-assisted control,'' {\em arXiv preprint arXiv:2302.07654}, 2023.

\bibitem{lehna2023managing}
M.~Lehna, J.~Viebahn, C.~Scholz, A.~Marot, and S.~Tomforde, ``Managing power grids through topology actions: A comparative study between advanced rule-based and reinforcement learning agents,'' 2023.

\bibitem{subramanian2021exploring}
M.~Subramanian, J.~Viebahn, S.~H. Tindemans, B.~Donnot, and A.~Marot, ``Exploring grid topology reconfiguration using a simple deep reinforcement learning approach,'' in {\em 2021 IEEE Madrid PowerTech}, pp.~1--6, IEEE, 2021.

\bibitem{aluisio2023remedial}
B.~Aluisio, M.~Barbetta, C.~Coluzzi, F.~M. Gatta, A.~Geri, S.~Lauria, M.~Maccioni, L.~Nati, and L.~Ortolano, ``A remedial action optimizer for the flow-based capacity calculation,'' in {\em 2023 AEIT International Annual Conference (AEIT)}, pp.~1--6, IEEE, 2023.

\bibitem{crognier2021grid}
G.~Crognier, P.~Tournebise, M.~Ruiz, and P.~Panciatici, ``Grid operation-based outage maintenance planning,'' {\em Electric Power Systems Research}, vol.~190, p.~106682, 2021.

\bibitem{dorfer2022power}
M.~Dorfer, A.~R. Fuxj{\"a}ger, K.~Kozak, P.~M. Blies, and M.~Wasserer, ``Power grid congestion management via topology optimization with alphazero,'' {\em arXiv preprint arXiv:2211.05612}, 2022.

\bibitem{Numan2023}
M.~Numan, M.~F. Abbas, M.~Yousif, S.~S. Ghoneim, A.~Mohammad, and A.~Noorwali, ``The role of optimal transmission switching in enhancing grid flexibility: A review,'' 2023.

\bibitem{ruiz2016security}
P.~A. Ruiz, E.~Goldis, A.~M. Rudkevich, M.~C. Caramanis, C.~R. Philbrick, and J.~M. Foster, ``Security-constrained transmission topology control milp formulation using sensitivity factors,'' {\em IEEE Transactions on Power Systems}, vol.~32, no.~2, pp.~1597--1605, 2016.

\bibitem{dehghanian2017power}
P.~Dehghanian, {\em Power system topology control for enhanced resilience of smart electricity grids}.
\newblock PhD thesis, 2017.

\bibitem{morsy2022security}
B.~Morsy, A.~Hinneck, D.~Pozo, and J.~Bialek, ``Security constrained opf utilizing substation reconfiguration and busbar splitting,'' {\em Electric Power Systems Research}, vol.~212, p.~108507, 2022.

\bibitem{Ruiz2017}
P.~A. Ruiz, E.~Goldis, A.~M. Rudkevich, M.~C. Caramanis, C.~R. Philbrick, and J.~M. Foster, ``Security-constrained transmission topology control milp formulation using sensitivity factors,'' {\em IEEE Transactions on Power Systems}, vol.~32, pp.~1597--1605, 3 2017.

\bibitem{Goldis2017}
E.~A. Goldis, P.~A. Ruiz, M.~C. Caramanis, X.~Li, C.~R. Philbrick, and A.~M. Rudkevich, ``Shift factor-based scopf topology control mip formulations with substation configurations,'' {\em IEEE Transactions on Power Systems}, vol.~32, pp.~1179--1190, 3 2017.

\bibitem{glavitsch1985switching}
H.~Glavitsch, ``Switching as means of control in the power system,'' {\em International Journal of Electrical Power \& Energy Systems}, vol.~7, no.~2, pp.~92--100, 1985.

\bibitem{marot2020learning}
A.~Marot, B.~Donnot, C.~Romero, B.~Donon, M.~Lerousseau, L.~Veyrin-Forrer, and I.~Guyon, ``Learning to run a power network challenge for training topology controllers,'' {\em Electric Power Systems Research}, vol.~189, p.~106635, 2020.

\bibitem{viebahn2022potential}
J.~Viebahn, M.~Naglic, A.~Marot, B.~Donnot, and S.~H. Tindemans, ``Potential and challenges of ai-powered decision support for short-term system operations,'' {\em CIGRE Session 2022}, 2022.

\bibitem{Sauer2001}
P.~W. Sauer, K.~E. Reinhard, and T.~J. Overbye, ``Hicss'01: Extended factors for linear contingency analysis,'' 2001.

\bibitem{marot2018expert}
A.~Marot, B.~Donnot, S.~Tazi, and P.~Panciatici, ``Expert system for topological remedial action discovery in smart grids,'' in {\em Mediterranean Conference on Power Generation, Transmission, Distribution and Energy Conversion (MEDPOWER 2018)}, pp.~1--6, IET, 2018.

\bibitem{viebahn2024potential}
J.~Viebahn, S.~Kop, J.~van DIJK, and al., ``Gridoptions tool: Real-world day-ahead congestion management using topological remedial actions,'' {\em CIGRE Session 2024}, 2024.

\bibitem{schnyder1990security}
G.~Schnyder and H.~Glavitsch, ``Security enhancement using an optimal switching power flow,'' {\em IEEE Transactions on Power Systems}, vol.~5, no.~2, pp.~674--681, 1990.

\bibitem{hedman2011review}
K.~W. Hedman, S.~S. Oren, and R.~P. O'Neill, ``A review of transmission switching and network topology optimization,'' in {\em 2011 IEEE power and energy society general meeting}, pp.~1--7, IEEE, 2011.

\bibitem{shahidehpour2003market}
M.~Shahidehpour, H.~Yamin, and Z.~Li, {\em Market operations in electric power systems: forecasting, scheduling, and risk management}.
\newblock John Wiley \& Sons, 2003.

\bibitem{wood2013power}
A.~J. Wood, B.~F. Wollenberg, and G.~B. Shebl{\'e}, {\em Power generation, operation, and control}.
\newblock John Wiley \& Sons, 2013.

\bibitem{bhattacharya2012operation}
K.~Bhattacharya, M.~H. Bollen, and J.~E. Daalder, {\em Operation of restructured power systems}.
\newblock Springer Science \& Business Media, 2012.

\bibitem{guler2007generalized}
T.~Guler, G.~Gross, and M.~Liu, ``Generalized line outage distribution factors,'' {\em IEEE Transactions on Power systems}, vol.~22, no.~2, pp.~879--881, 2007.

\bibitem{van2023bus}
J.~van Dijk, J.~Viebahn, B.~Cijsouw, and J.~van Casteren, ``Bus split distribution factors,'' {\em IEEE Transactions on Power Systems}, 2023.

\bibitem{zaoui2005coupling}
F.~Zaoui, S.~Fliscounakis, and R.~Gonzalez, ``Coupling opf and topology optimization for security purposes,'' in {\em 15th Power Systems Computation Conference}, pp.~22--26, 2005.

\bibitem{fliscounakis2007topology}
S.~Fliscounakis, F.~Zaoui, G.~Sim{\'e}ant, and R.~Gonzalez, ``Topology influence on loss reduction as a mixed integer linear programming problem,'' in {\em 2007 IEEE Lausanne Power Tech}, pp.~1987--1990, IEEE, 2007.

\bibitem{Heidarifar2016}
M.~Heidarifar and H.~Ghasemi, ``A network topology optimization model based on substation and node-breaker modeling,'' {\em IEEE Transactions on Power Systems}, vol.~31, pp.~247--255, 1 2016.

\bibitem{boylestad2009electronic}
R.~L. Boylestad and L.~Nashelsky, {\em Electronic devices and circuit theory}.
\newblock Pearson Education India, 2009.

\bibitem{stott2009dc}
B.~Stott, J.~Jardim, and O.~Alsa{\c{c}}, ``Dc power flow revisited,'' {\em IEEE Transactions on Power Systems}, vol.~24, no.~3, pp.~1290--1300, 2009.

\bibitem{goldis2016shift}
E.~A. Goldis, P.~A. Ruiz, M.~C. Caramanis, X.~Li, C.~R. Philbrick, and A.~M. Rudkevich, ``Shift factor-based scopf topology control mip formulations with substation configurations,'' {\em IEEE Transactions on Power Systems}, vol.~32, no.~2, pp.~1179--1190, 2016.

\bibitem{grid2op}
B.~Donnot, ``{Grid2op- A testbed platform to model sequential decision making in power systems. }.'' \url{https://GitHub.com/rte-france/grid2op}, 2020.

\bibitem{lightsim2grid}
B.~Donnot, ``{Lightsim2grid - A c++ backend targeting the Grid2Op platform. }.'' \url{https://GitHub.com/bdonnot/lightsim2grid}, 2020.

\bibitem{thurner2018pandapower}
L.~Thurner, A.~Scheidler, F.~Sch{\"a}fer, J.-H. Menke, J.~Dollichon, F.~Meier, S.~Meinecke, and M.~Braun, ``pandapower—an open-source python tool for convenient modeling, analysis, and optimization of electric power systems,'' {\em IEEE Transactions on Power Systems}, vol.~33, no.~6, pp.~6510--6521, 2018.

\bibitem{Marot2018Guided}
A.~{Marot}, S.~{Tazi}, B.~{Donnot}, and P.~{Panciatici}, ``Guided machine learning for power grid segmentation,'' in {\em 2018 IEEE PES Innovative Smart Grid Technologies Conference Europe (ISGT-Europe)}, pp.~1--6, 2018.

\bibitem{HENKA2022Segmentation}
N.~Henka, Q.~Francois, S.~Tazi, M.~Ruiz, and P.~Panciatici, ``Power grid segmentation for local topological controllers,'' {\em Electric Power Systems Research}, vol.~213, p.~108302, 2022.

\bibitem{donon2020leap}
B.~Donon, B.~Donnot, I.~Guyon, Z.~Liu, A.~Marot, P.~Panciatici, and M.~Schoenauer, ``Leap nets for system identification and application to power systems,'' {\em Neurocomputing}, vol.~416, pp.~316--327, 2020.

\end{thebibliography}

\printnomenclature

\appendix

\subsection{Multiple lines reconnection recursive demonstration}
\label{appendix:multi_reco}
After the two-line reconnection case in section \ref{sec:line_reco}, we now consider the reconnection of three lines in $R=\{l_1,l_2,l_3\}$ from a topology $\Tau_O$ in which the three are disconnected to reach a topology $\Tau_R$ in which they are all reconnected. We also consider the subsets of two reconnected lines defined as $R_{\setminus l_i}$ for which the third line $l_i$ remains disconnected. EST coefficients $\alpha_R$ and $\beta^{l_i}_R$ for superset $R$ will be derived as a recursive formula from previous rank subsets $R_{\setminus l_i}$. 

We start by reusing EST for line disconnections as previously:
\begin{equation}
\label{eqn:sup_topo_theorem_reco_3_lines_init}
    PF(\Tau_{O}) = \alpha_{O} PF(\Tau_{R}) +  \sum_{l_i \in R}
    \beta^{l_i}_{O} PF(\Tau^{\tau_{\setminus l_i}}_{R}) 
\end{equation}

Applying (\ref{eqn:sup_topo_theorem_reco_2_lines}) for two line reconnections, since $\Tau^{\tau_{\setminus l_1}}_{R}=\Tau_{R}\circ \tau_{\setminus l_1}=\Tau_{O}\circ \tau_{=l_2}\circ \tau_{=l_3} = \Tau^{\tau_{=\{l_2,l_3\}}}_{O}$, we now have here :

\begin{equation}
\label{eqn:sup_topo_theorem_reco_3_lines_2_lines}
    \begin{split}
        PF(\Tau^{\tau_{\setminus l_1}}_{R}) & = PF(\Tau^{\tau_{=\{l_2,l_3\}}}_{O})\\
        & = \alpha_{R_{\setminus l_1}} PF(\Tau_{O}) + \sum_{l_j \in R_{\setminus l_1}} \beta_{R_{\setminus l_1}}^{l_j} PF(\Tau^{\tau_{=l_j}}_{O})
    \end{split}
\end{equation}

Transposing (\ref{eqn:sup_topo_theorem_reco_3_lines_2_lines}) to each $\Tau^{\tau_{\setminus l_i}}$ for $i \in {1,2,3}$ and substituting it in (\ref{eqn:sup_topo_theorem_reco_3_lines_init}) we obtain:

\begin{multline}
    PF(\Tau_{O}) = \alpha_{O} PF(\Tau_{R}) \\
    +  \sum_{l_i \in R} \beta^{l_i}_{O} (\alpha_{R_{\setminus l_i}} PF(\Tau_{O}) + \sum_{l_j \in R_{\setminus l_i}} \beta_{R_{\setminus l_i}}^{l_j} PF(\Tau^{\tau_{=l_j}}_{O}))
\end{multline}

To finally reach: 
\begin{multline}
    PF(\Tau_{R})=\frac{1}{\alpha_{O}}((1- \sum_{l_i \in R} \beta^{l_i}_{O} (\alpha_{R_{\setminus l_i}}))PF(\Tau_{O}) + \\
    (-\beta^{l_i}_{O}\sum_{l_j \in R_{\setminus l_i}} \beta_{R_{\setminus l_i}}^{l_j})PF(\Tau^{\tau_{=l_j}}_{O}))
\end{multline}

We recover the ST equation:
\begin{multline}
\label{eqn:sup_topo_theorem_reco_3_lines_final}
    PF(\Tau_{R}) = \alpha_{R} PF(\Tau_{O}) + \sum_{l_i \in R} \beta_{R}^{l_i} PF(\Tau^{\tau_{=l_i}}_{O})\\
    PF(\Tau^{tgt}) = \alpha_{R} PF(\Tau^{ref}) + \sum_{l_i \in R} \beta_{R}^{l_i} PF(\Tau^{\tau_{=l_i}})\\
    \text{with } \alpha_{R}=\frac{1-\sum_{l_i \in R} \beta^{l_i}_{O} \alpha_{R_{\setminus l_i}}}{\alpha_{O}} \\ \text{ and } \beta_{R}^{l_i}=\frac{-\beta^{l_i}_{O}\sum_{l_j \in R_{\setminus l_i}} \beta_{R_{\setminus l_i}}^{l_j}}{\alpha_{O}}\\
\end{multline}

Again, this works out when $\alpha_{O}$ is non-null. As $\alpha_R$ and $\beta^{l_i}$ are all properly defined from (\ref{eqn:sup_topo_theorem_reco_beta_value}) and cannot be infinite, $\alpha_{O}$ is in fact never null.

Reusing (\ref{eqn:alphas_2_lines_reco}) we have:
\begin{equation}
    \alpha_{R_{\setminus l_i}} = 1 - \sum_{j \in \{1,2,3\}, j \neq i} \beta_{R_{\setminus l_i}}^j
\end{equation}

Which we substitutes in (\ref{eqn:sup_topo_theorem_reco_3_lines_final}) to recover:
\begin{equation}
\label{eqn:alphas_3_lines_reco}
    \alpha_{R} = 1 - \sum_{i \in \{1,2,3\}} \beta^{l_i}_{R}
\end{equation}

The same recursion applies for cases of more than 3 line disconnections, as all above equations hold for the next recursive ranks. 
\subsection{Injected canceling flows cancel out}
\label{appendix:cancel_out}

We here demonstrate that unknown injected canceling flows $cf_l$ cancels out in the EST equation as required in section \ref{sec:eqn_system_reconnection}. 

In the equivalent model of canceling flows (\ref{eqn:virtual_flow_model}), considering the base topology with lines reconnected in $\Tau^{tgt}=\Tau_R$, equation (\ref{eqn:sup_topo_theorem_all_reco}) can be rewritten as:

\begin{multline}
\label{eqn:sup_th_reco_injected_flow}
    PF(P_{g-d}^{ref},\Tau_R) = \alpha_{R} PF(P_{g-d}^{ref}+Cf^{T_O},\Tau_R)\\
    + \sum_{l_i \in R} \beta_R^{l_i}  PF(P_{g-d}^{ref}+Cf^{T_O^{\tau_{=l_i}}},\Tau_R)
\end{multline}

Note the canceling flow additions on the right-hand sides to make equivalent power flows in this base topology $T_R$ compared to the initially different topologies $T_O$ and $T_O^{\tau_{=l_i}}$. 

Using the standard ST, we further have:
\begin{equation}
 PF(P_{g-d}^{ref}+Cf^{T_O}) = PF(P_{g-d}^{ref}) + PF(Cf^{T_O})
\end{equation}

So we can rearrange (\ref{eqn:sup_th_reco_injected_flow}):
\begin{multline}
    (1- \alpha_{R} - \sum_{l \in R} \beta_R^{l_i})PF(P_{g-d}^{ref}) =  PF (\alpha_{R} Cf^{T_O}) \\ +  \sum_{l \in R} PF(\beta_R^{l_i} Cf^{T_O^{\tau_{=l_i}}})  
\end{multline}

with left hand side null given (\ref{eqn:alphas_3_lines_reco}):
\begin{equation}
\label{eqn:sup_th_reco_injected_flow_2}
    (1- \alpha_{R} - \sum_{l \in R} \beta_R^{l_i}) = 0   
\end{equation}

So right hand side is also null:
\begin{multline}
\begin{split}
 PF (\alpha_{R} Cf^{T_O}) +  \sum_{l \in R} PF(\beta_R^{l_i} Cf^{T_O^{\tau_{=l_i}}}) & = 0\\
    PF (\alpha_{R} Cf^{T_O} + \sum_{l \in R} \beta_R^{l_i} Cf^{T_O^{\tau_{=l_i}}}) & = 0     
\end{split}
\end{multline}

Null power flows all over the grid is only possible if all nodal power balance are null. This leads to:
\begin{multline}
\begin{split}
\alpha_{R} Cf^{T_O} + \sum_{l \in R} \beta_R^{l_i} Cf^{T_O^{\tau_{=l_i}}} & = 0\\
\alpha_R cf^{\Tau_O}_{l} + \sum_{l_i \in R_{\setminus l} }\beta_R^{l_i} cf^{T_O^{\tau_{=l_i}}}_{l}& = 0 \text{ for all $l \in R$ as $cf^{T_O^{\tau_{=l}}}_{l}=0$} 
\end{split}
\end{multline}

\end{document}